\def\year{2015}
\documentclass[letterpaper]{article}
\usepackage{aaai}
\usepackage{times}
\usepackage{helvet}
\usepackage{courier}
\usepackage[utf8]{inputenc}
\usepackage[english]{babel}

\usepackage{amssymb}
\usepackage{graphicx}
\usepackage{url}
\usepackage{tabularx}
\usepackage{multicol}
\usepackage{multirow}
\usepackage{xspace}
\usepackage{booktabs}
\usepackage{subfigure}
\usepackage{amsmath,graphicx} 
\usepackage{color}
\usepackage{comment}

\excludecomment{comment}

\fontfamily{pbk}\selectfont

\usepackage{footnote}
\usepackage{titlesec}

\makesavenoteenv{tabular}
\makesavenoteenv{table}

\begin{document}

\date{}

\title{It's a Man's Wikipedia? \\Assessing Gender Inequality in an Online Encyclopedia}  

\urldef{\mailgesis}\path|claudia.wagner, mohsen.jadidi, markus.strohmaier@gesis.org|
\urldef{\maileth}\path|dgarcia@ethz.ch|

\author{
  Claudia Wagner, David Garcia, Mohsen Jadidi, Markus Strohmaier\\
  GESIS - Leibniz Institute for the Social Sciences, ETH Zürich,  University of Koblenz-Landau  \\
  \mailgesis \\  \maileth 
}

\author{Claudia Wagner  \\
	GESIS \& Uni. Koblenz \\
	claudia.wagner@gesis.org \\
	\And
	 David Garcia \\
	ETH Zürich \\
	dgarcia@ethz.ch \\
	\And 
	 Mohsen Jadidi \\
	 GESIS \\
	  mohsen.jadidi@gesis.org \\
	  \And 
	  Markus Strohmaier \\
	  GESIS \& Uni. Koblenz \\
	  markus.strohmaier@gesis.org
	}

\maketitle




\urldef{\TFA}\url{http://en.wikipedia.org/wiki/Wikipedia:Today%27s_featured_articlef}

\thispagestyle{empty}

\begin{abstract}
Wikipedia is a community-created encyclopedia that contains information about notable people from different countries, epochs and disciplines and aims to document the world's knowledge from a neutral point of view.
However, the narrow diversity of the Wikipedia editor community 
has the potential to introduce systemic biases such as gender biases into the content of Wikipedia.
In this paper we aim to tackle a sub problem of this larger challenge by presenting and applying a computational method for \emph{assessing gender bias on Wikipedia along multiple dimensions}. We find that while women on Wikipedia are covered and featured well in many Wikipedia language editions, \emph{the way women are portrayed} starkly differs from the way men are portrayed. We hope our work contributes to increasing awareness about gender biases online, and in particular to raising attention to the different levels in which gender biases can manifest themselves on the web. 
\end{abstract}



\section{Introduction}

Wikipedia aims to provide a platform to freely share the sum of all human knowledge. It represents an influential source of information on the web, containing encyclopedic information about notable people from different countries, epochs and disciplines that is used for learning and educational purposes worldwide. Wikipedia is also a community-created effort driven by a self-selected set of editors. The demographic characteristics of this set of editors is known: it is predominately white and male \cite{LamUDSMTR11,Collier2012,Hill2013}. 

This known gender bias in the population of editors has the potential to introduce gender biases into the contents of Wikipedia as well. For example, the population bias might lead to differences in the ways women and men are portrayed on Wikipedia. It might also mimic or even exaggerate inequalities that are already existing in the real world. At the same time, assessing the manifold and subtle ways in which gender biases can manifest themselves has been challenging, and we know little about the different dimensions of gender biases on Wikipedia. Yet, due to the influential nature of Wikipedia, it is important to reveal, assess and correct such biases, if they exist. This paper tackles a sub-part of this larger challenge.

\textbf{Objectives:} In particular, the overall goal of this work is to \emph{assess potential gender inequalities in Wikipedia articles} along different dimensions. 

\textbf{Approach:} To assess the extent to which Wikipedia suffers from potential gender bias, we analyze articles about notable people in six language editions along four different gender bias dimensions: coverage bias, structural bias,  lexical bias and visibility bias. 
\emph{Coverage bias} determines differences between the number of notable women and men portrayed on Wikipedia. 
For example, one might hypothesize that notable men are more likely to be covered by Wikipedia. 
\emph{Structural bias} quantifies gender homophily/disassortativity, i.e. gender-specific tendencies to preferably link articles of notable people with the same or different gender. 
For example, one might hypothesise that articles about women have more links to men than vice versa. 
\emph{Lexical bias} reveals inequalities in the words used to describe notable men and women on Wikipedia. 
For example, articles about women are potentially more likely to mention their family (husband or kids) than articles about men. 
\emph{Visibility bias} reflects how many articles about men or women make it to the front page of Wikipedia. Again, one can hypothesize that articles about men might have better chances to be selected.


\textbf{Contributions \& Findings:} We present and apply a computational method for \emph{assessing gender bias on Wikipedia along multiple dimensions}. 
We find that most Wikipedia language editions exhibit a slight over-representation of women, but the proportional differences in the coverage of men and women are not significant. That means, men and women are covered equally well in all six Wikipedia language editions.
Also on the visibility level, we do not find any evidence for male-bias in the selection procedure of articles that are featured on the startpage of the English  Wikipedia.
These are encouraging findings suggesting that the Wikipedia editor community is sensible to gender inequalities\footnote{also cf. \url{http://meta.wikimedia.org/wiki/Gender_gap}} and covers notable women and men equally well.
However, we also find that \emph{the way women are portrayed} on Wikipedia starkly differs from the way men are portrayed. We find evidence for both structural and lexical gender biases. On a structural level, we observe an asymmetry: Women on Wikipedia tend to be more linked to men than vice versa. On a lexical level we find that especially romantic relationships and family-related issues are much more frequently discussed on Wikipedia articles about women than men.

\begin{savenotes}
\begin{table}[b!]
\centering
\footnotesize
\caption{\textbf{Statistics of the datasets:} The number of articles and median article length of all Wikipedia articles that belong to one of the notable people from our three reference datasets.}
\begin{tabular}{p{3cm}|lll}
& Freebase & HA & Pantheon\\ \hline
Total Num Articles &109,481\ & 4,002  & 11,341\\
Female Articles &12,685 & 88 & 1,496\\
Male Articles & 96,796 & 3,914 & 9,845\\ 
\hline
Median Num Words Female & 458 & 1,121 & 1,106 \\
Median Num Words Male &  412 &  820 & 1,017 \\
 \hline

\end{tabular}
\label{tab:datasets_desc}
\end{table}
\end{savenotes}


\begin{figure}[t!]
\centering
\includegraphics[width=0.55\linewidth]{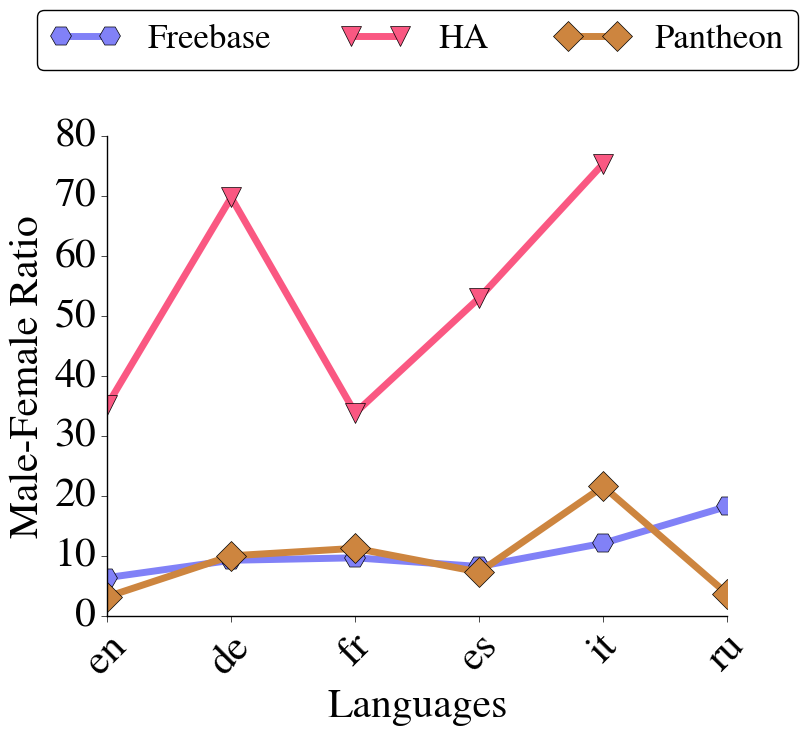}
\caption{\textbf{Male-Female Ratio:} The ratio of men and women in our reference datasets that are born in a country where one of the six languages is predominantly spoken. Across all language editions the local heroes of a country tend to be predominantly male. For example, if we look at notable people in freebase we find between 7 and 12 times more men than women depending on which countries we consider.}
\label{fig:birthplace}
\end{figure}

\section{Materials \& Methods}
\label{sec:methods}
In the following we discuss our data collection and our methodology that allows to systematically explore gender inequalities on Wikipedia on multiple dimensions. 

\subsection{Datasets}

To estimate the bias on Wikipedia that goes beyond the bias in the offline world, ideally one would have a complete list of notable people available that is (a) not biased and (b) independent from Wikipedia. Since it is impossible to obtain such a list, we use the following three collections of notable people as \emph{reference datasets}, each having different strength and weaknesses:

\textbf{Freebase}: We use a collection of around 120k notable people that has been used in previous research for studying the mobility of notable people \cite{Schich2014} and was obtained from freebase. 
Freebase contains data harvested from sources such as Wikipedia, NNDB, FMD and MusicBrainz, as well as individually contributed data from users.
We only take individuals into account for which gender and basic bibliographic information (i.e.,full birth and 
death date and birth and death location) is available. 
Freebase directly links to Wikipedia articles in different language editions, if articles about the entity are available.  


\textbf{Pantheon}: Pantheon is a project developed by the Macro Connections group at the MIT Media Lab that is collecting, analyzing, and visualizing data on historical cultural popularity and production.
The Pantheon dataset \cite{Pantheon} contains information on 11,340 biographies that have presence in more than 25 languages in the Wikipedia (as of May 2013) and provides links to Wikipedia articles about these people.


\textbf{Human Accomplishment}: The third dataset which we use is compiled from a book called ``Human Accomplishment'' \cite{Murray2003} (short HA) and contains information on 4,002 eminent individuals from arts and sciences who made a significant contribution prior to 1950. The inventories were constructed by Charles Murray using linguistic records, such as encyclopedia entries from a number of different languages and sources.
Also this dataset has biases since e.g. Murray relied mainly on materials in Roman-alphabet languages.
To find Wikipedia articles about those individuals, we use the Wikipedia search API and search for the full name. To select the right search result from the list we compare the birth date, birth location, death date and death location of the candidates in the search results with the person we are looking for.  



\begin{figure*}[t!]
  \centering
\subfigure[Freebase]{
\includegraphics[width=0.2\linewidth]{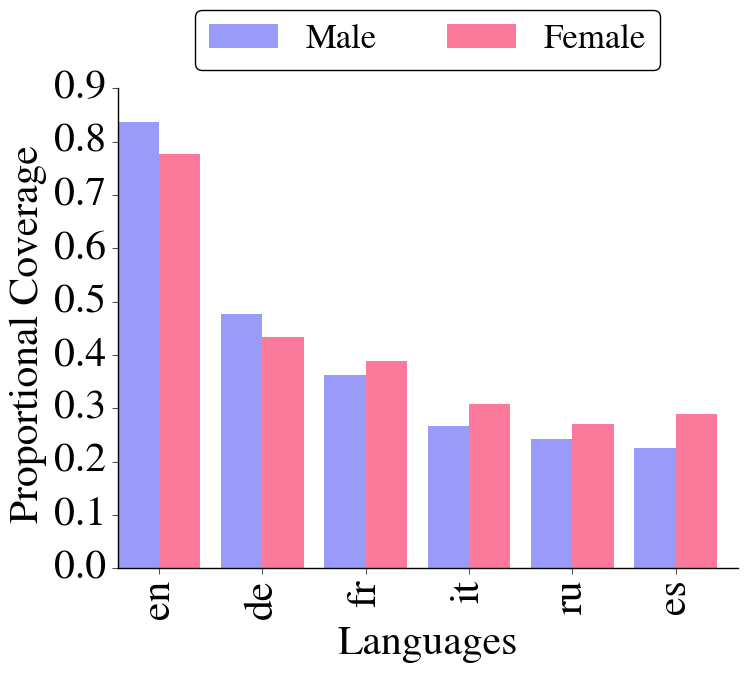}
}
\subfigure[HA]{
  \includegraphics[width=0.2\linewidth]{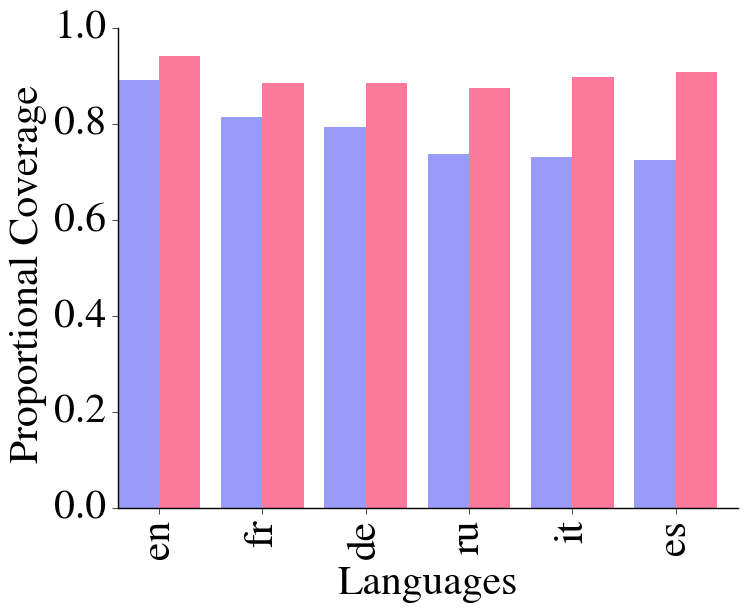}
} 
\subfigure[Pantheon]{
  \includegraphics[width=0.2\linewidth]{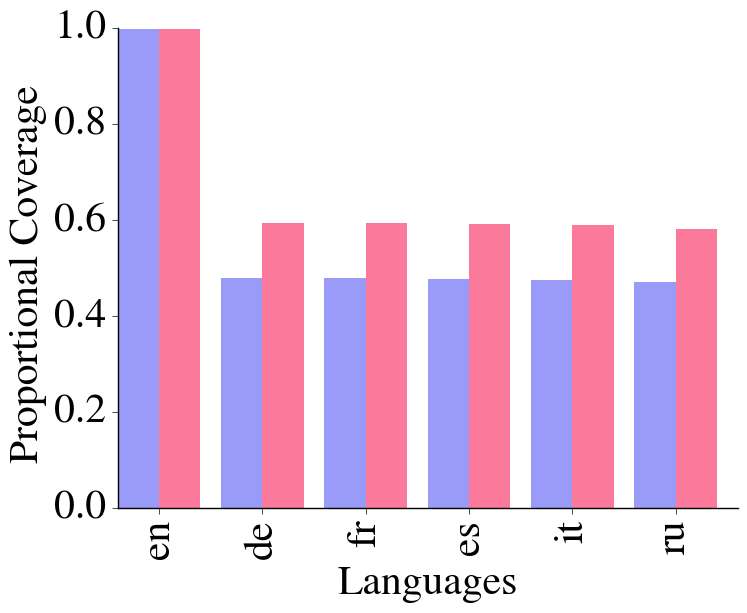}
} 
\caption{\textbf{Coverage Bias:} Proportional coverage of notable women and men. 
Surprisingly, in most language editions the proportion of notable women covered is slightly higher than the proportion of notable men. 
 }
\label{fig:prop-selected} 
\end{figure*}

\textbf{Data Collection Procedure}:
We crawled the content of articles about people in our reference datasets using Wikipedia's API in November 2014. 
For the English Wikipedia, the articles that have been featured at the front page in the last few years were extracted from the ``Today's Featured Article'' archive\footnote{\TFA}.
Table~\ref{tab:datasets_desc} provides the basic statistics for each dataset and Figure \ref{fig:birthplace} shows the ratio between men and women that are born in a country where one of the six languages we studied is predominantly spoken. 
The overlap between the three reference datasets is very low. For example, for those people from our reference datasets which we could map to the English Wikipedia the Jaccard coefficient is $0.016$ for freebase and HA, $0.035$ for freebase and pantheon and $0.097$ for pantheon and HA.
The six language editions that we explore in this study are those which had the highest coverage of notable men and women from our largest reference dataset, freebase.

\subsection{Measuring Gender Inequality}

We propose to analyze gender inequality on Wikipedia
 on the following four dimensions: which notable men or women are presented on Wikipedia (coverage bias)? How are they presented (lexical bias)? What structure emerges from the hyperlink network of articles (structural bias)? And which articles get featured on the startpage of Wikipedia (visibility bias)?

\textbf{Coverage Bias}: To estimate coverage bias we compare the proportions of notable men and women of different reference datasets that are covered by Wikipedia.
Ideally, a reference dataset consists of an unbiased list of people who should be presented on Wikipedia.
It is important to understand that a biased reference dataset will obviously impact our results.
If, for example, our reference dataset is already biased towards men (i.e., it covers only extremely famous women but also less famous men) than the proportion of women who are represented on Wikipedia would probably be higher than the proportion of men.
To address this issue we analyze the coverage using several independent reference datasets (Jaccard coefficient between the three datasets ranges from $0.0$ to $0.12$ for different language editions), assuming that each of them will have a different bias and seeking patterns that exist across all three datasets.


Further, gender-differences in the extent to which men and women are covered on Wikipedia may exist. Therefore, we also analyse the article length distribution of men and women. 

\begin{figure}[t!]
  \centering
  \includegraphics[width=0.6\linewidth]{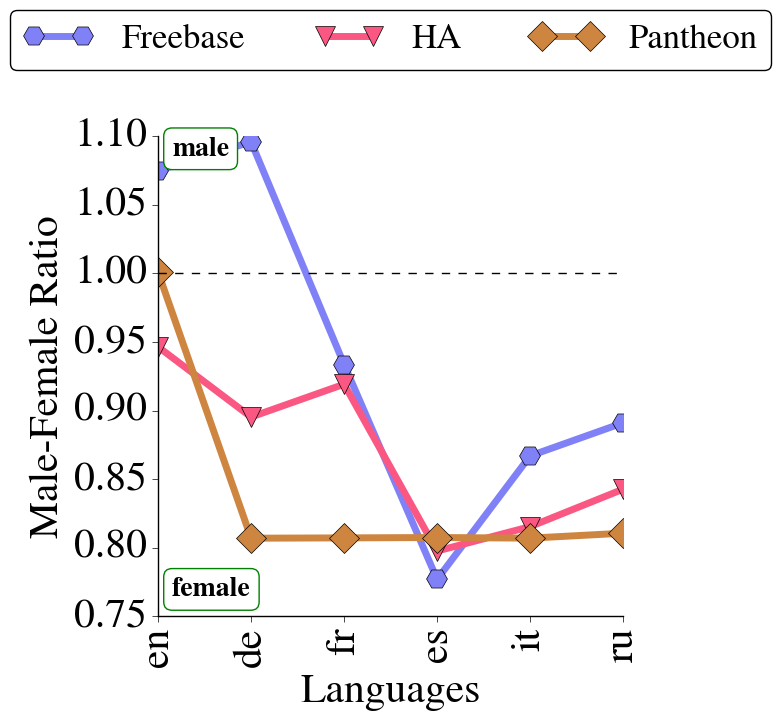}
\caption{\textbf{Coverage Gap:} Ratio between the number of notable men and women from three different reference lists that are covered on different language editions of Wikipedia. 
 }
\label{fig:prop-gap} 
\end{figure}



\textbf{Structural Bias}: We analyze the patterns of gender assortativity based on the probability that an article about a person of one gender links to an article about a person of the other gender. We compare the probability that a link ends in an article of gender $g_2$ given that it comes from an article of gender $g_1$  with the probability that a link ends in an article of gender $g_2$ regardless of the gender of its origin:
\begin{equation}
L(g_1,g_2) = log\left (\frac{P(to=g_2 | from=g_1)}{P(to=g_2)}\right )   \label{eq:L}
\end{equation}
where $P(to=g_2 | from=g_1)$ is the conditional distribution that an edge links to an article of gender $g_2$ given that it comes from an article of gender $g_1$, and $P(to=g_2)$ is the probability that any link ends in an article of gender $g_2$ regardless of the gender of its origin. $L$ measures the log likelihood ratio between edge probabilities, comparing the posterior probability of finding a gender at the edge of a link given that we know the gender of its origin, and comparing it with the base rate of linking to an article of gender $g_2$. This way, positive values of $L$ indicate increased connectivity from $g_1$ to $g_2$, and negative values the opposite, and define a c assortativity matrix of the four combinations of genders that measures the tendencies to connect within and across genders.

For the case of same gender connections we use the standard definition of assortativity \cite{Newman2003}:
\begin{equation}
\frac{
\sum_g P(from=g, to=g) - P(from=g)*P(to=g)}
{1-\sum_g P(from=g)*P(to=g)}
\label{eq:assortativity}
\end{equation}
For the case of asymmetry across genders, we compare the entries of $L$ from one gender to the other, as $A=L(F,M) - L(M,F)$. Positive values of $A$ will indicate a stronger tendency of articles about women to connect to articles about men than the opposite, controlling for the difference in in-degrees and sizes of both genders.

The finding of gender assortativity and asymmetry between genders requires a test that allows us to compare our empirical estimates against null models of the network. For that reason, we set up numerical simulations of three different null models: a \emph{randomized gender} model in which we shuffle the genders of nodes; a \emph{randomized link end} model in which we rewire links to random articles, maintaining out degrees but fully randomizing in-degree;
and a \emph{randomized link origin} model, in which we maintain link ends but rewire their origin to an article sampled at random, which maintains in-degrees but randomizes out degrees. We run each simulation 10,000 times, recording values of assortativity and asymmetry to measure the mean and 95\% confidence intervals of these two statistics under each null model.

Structural biases can also manifest in the centrality measures, as suggested by the \emph{Smurfette principle} \cite{Pollitt1991}. 
That means, women can be positioned in the periphery of a network with a core composed of men. In that case the centrality of women would be lower. We operationalize centrality on Wikipedia as a quantification of importance, measuring the in-degree and k-coreness of an article. The in-degree of article $p$ is trivially calculated as the amount of articles that link to article $p$, and the in k-coreness is computed through a pruning mechanism based on in-degree \cite{Giatsidis2013}.

\textbf{Lexical Bias}: To explore gender-specific lexical inequalities on Wikipedia we use an open vocabulary approach, inspired by \cite{Schwartz2013}.
An open-vocabulary approach is not limited to predefined word lists, but linguistics are automatically determined from the text.
We compute the tfidf scores of the word stems obtained from a Snowball Stemmer and use them as features to train a Naive Bayes classifier.
The classifier determines which words are most effective in distinguishing the gender of the person an article is about.
Log likelihood ratios $L(word,g)$ are used for comparing different feature-outcome relationships.
\begin{equation}
L(word,g) = log\left (\frac{P(word|g)}{P(word)}\right )   \label{eq:Lword}
\end{equation}
where $P(word|g)$ is the conditional distribution that a word shows up in an article about a person given that the person's  gender is $g$, and $P(word)$ is the probability that a word shows up in any article regardless of the gender of the person the article is about.

The \emph{Finkbeiner test} \cite{Finkbeiner}
suggests that articles about women often emphasize the fact that she is a woman, mention her husband and his job, her kids and child care arrangements, how she nurtures her underlings, how she was taken aback by the competitiveness in her field and how she is such a role model for other women.
Also the historian Gillian Thomas who investigated the role of women in Britannica states in her book \cite{Thomas1992} that 
as contributors, women were relegated to matters of ``social and purely feminine affairs'' and as subjects, women were often little more than
addenda to male biographies (e.g., Marie Curie as the wife of Pierre Curie).

We create the following three categories of words that capture some aspects that could be over-represented in articles about women according to what Thomas observed in the Britannica and what the Finkbeiner test suggest:
\begin{itemize}
\item \emph{Gender} category contains words that emphasize that someone is a man or woman (i.e., man, women, mrs, mrs, lady, gentleman) 
\item \emph{Relationship} category consists of words about romantic relationships (e.g., married, divorced, couple, husband, wife)
\item \emph{Family} category aggregates words about family relations (e.g., kids, children, mother, grandmother).
\end{itemize}
All other words that cannot be assigned to the above mentioned categories fall into the category \emph{Others}.
To gain further insights into the types of words that have the highest log likelihood ratio for articles about men or women, native speakers of each language manually code the 150 words which are most useful for differentiating articles about men and women in each language edition. 


\textbf{Visibility Bias}: To estimate visibility bias we simply compare the proportions of notable men and women of different reference datasets that got featured on the startpage of the English Wikipedia. We test the significance of the difference in proportions between men and women that got featured using a Chi-Square test.


\begin{figure}[t!]
\centering
\includegraphics[width=0.8\linewidth]{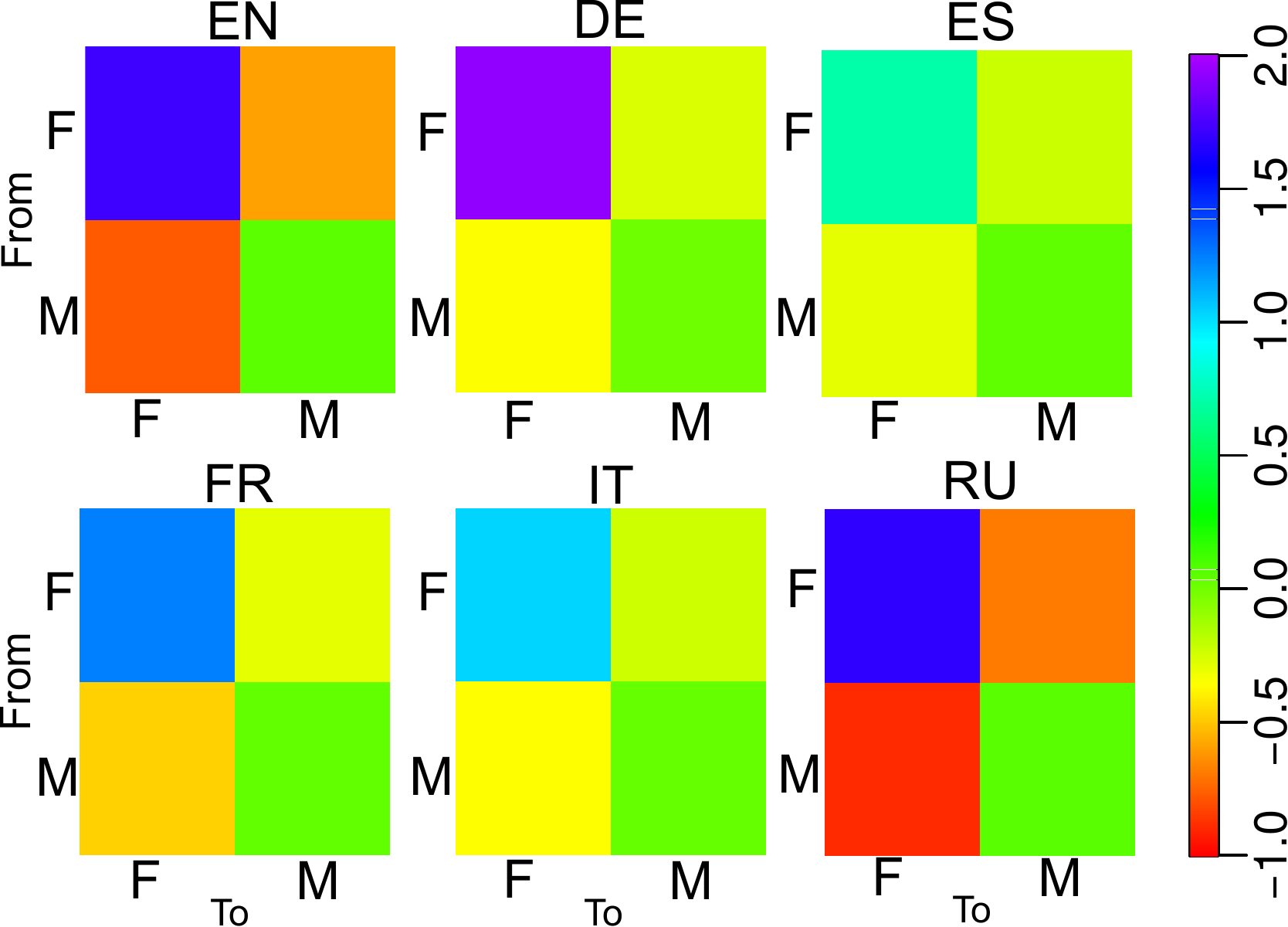}
\caption{\textbf{Structural Assortativity and Asymmetry Bias:} Logarithmic assortativity matrices for the hyperlink networks of articles about notable men and women in six language editions of Wikipedia. Assortativity of connections within genders becomes apparent for the minority class, women. All language editions show an asymmetry of connectivity across genders. The strongest assortativity and asymmetry is visible in the English and Russian Wikipedia.}
\label{fig:assortMatrices}
\end{figure}

\begin{figure}[t!]
\centering
\includegraphics[width=0.75\linewidth]{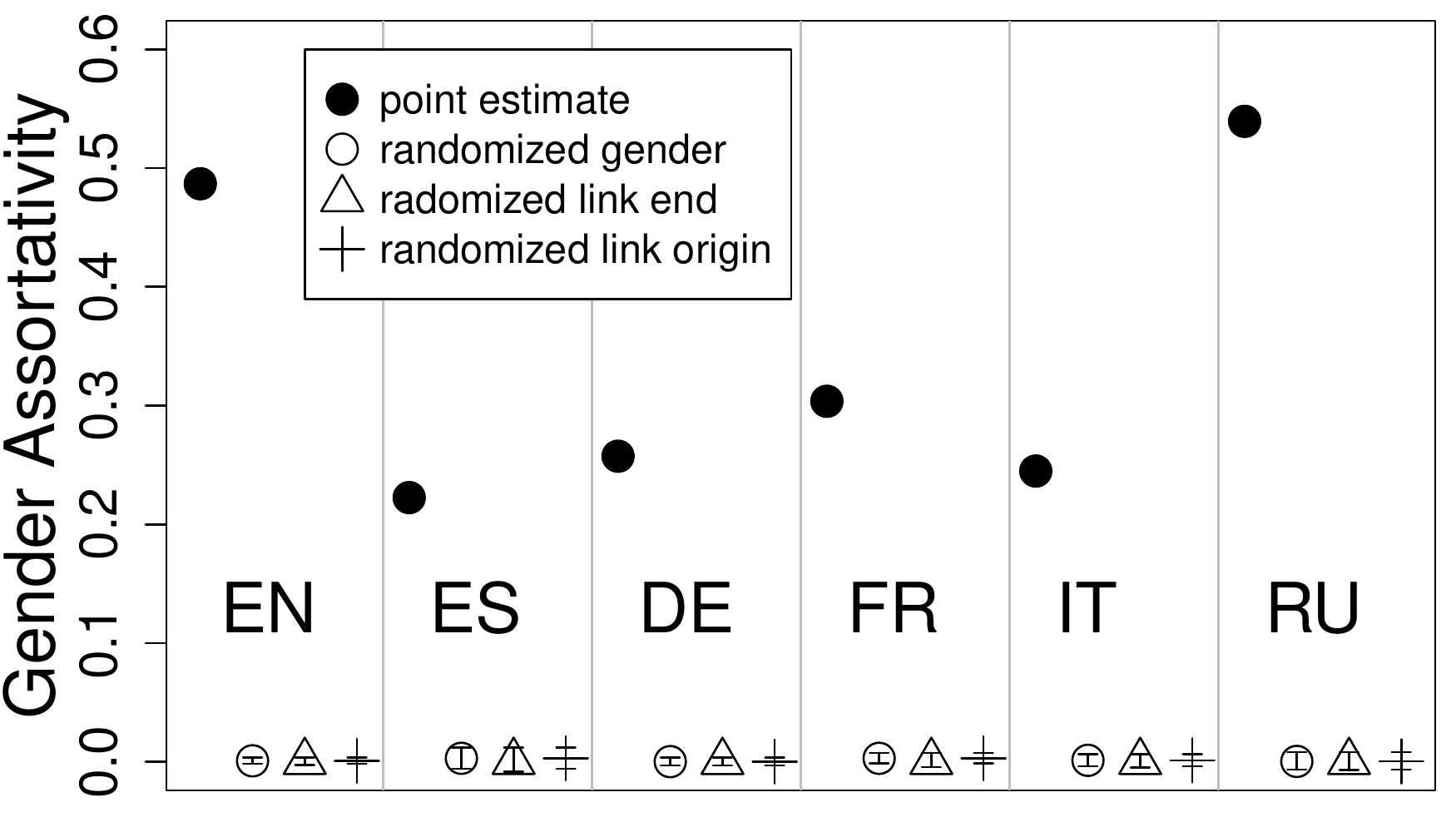}

\caption{\textbf{Significance of Structural Assortativity Bias:} Point estimates of gender assortativity in six language editions and comparison with the three null reference models. Error bars (smaller than symbol size) show 95\% confidence intervals over 10,000 simulations of each model. The empirical estimates
are significant in comparison to the narrow confidence interval of the null models.}
\label{fig:assort}
\end{figure}

\begin{figure}[t!]
\centering
 
\includegraphics[width=0.75\linewidth]{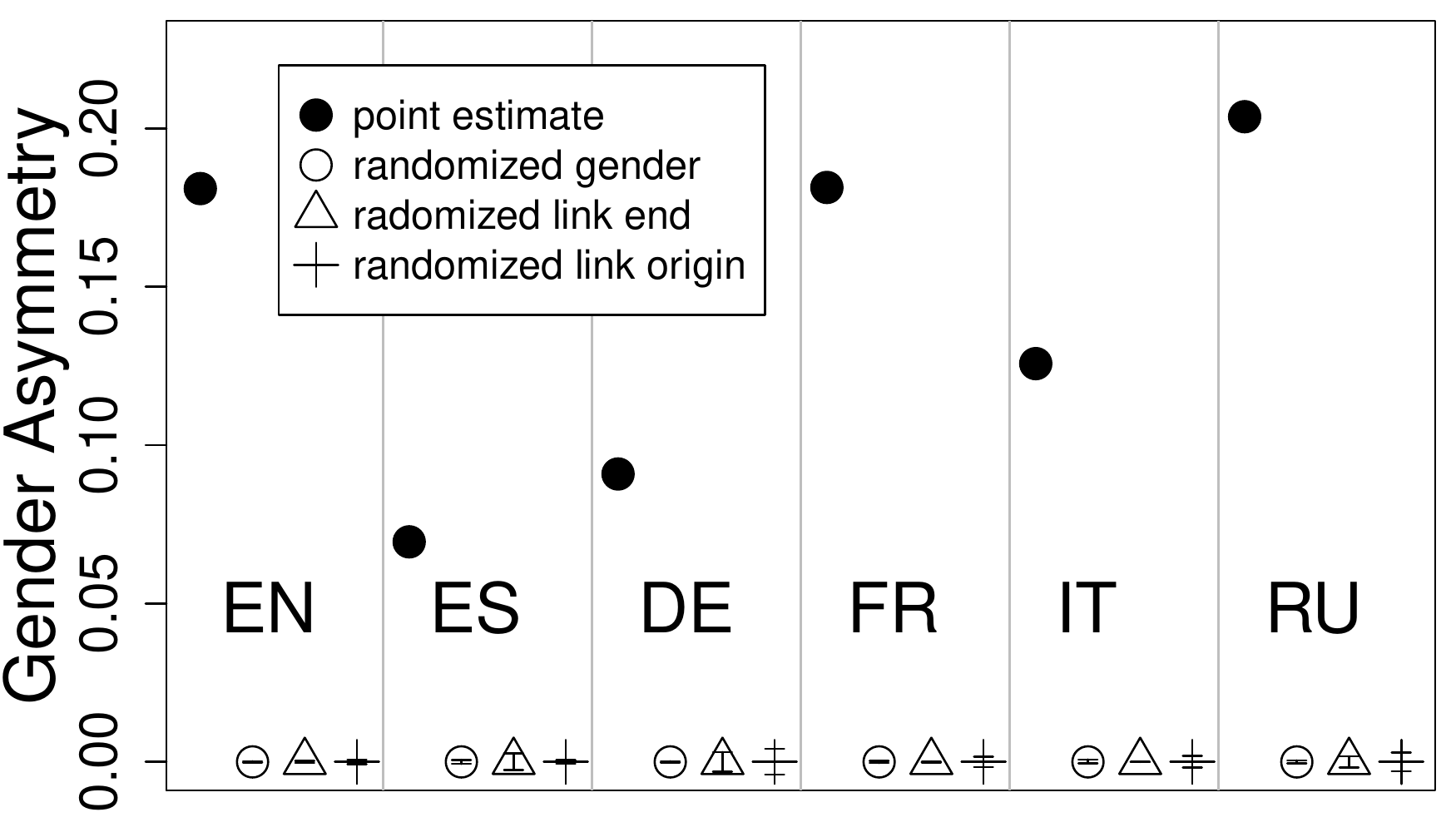}

\caption{\textbf{Significance of Structural Asymmetry Bias:} Arithmetic mean of point estimates of gender asymmetry for men and women in six language editions and comparison with the three null reference models. Error bars (smaller than symbol size) show 95\% confidence intervals over 10,000 simulations of each model. The empirical estimates
are significant in comparison to the narrow confidence interval of the null models.}
\label{fig:asymmetries}
\end{figure}

\section{Results}
\label{sec:results}
In the following, we present our empirical results on gender inequality on Wikipedia.


\subsection{Coverage Bias}
Figure \ref{fig:prop-selected}  shows that the best coverage across languages is achieved for people that made significant contributions to science and arts before 1950 and are therefore listed in the HA reference dataset.
Across all three reference datasets we consistently observe that \emph{women are} not - as initially hypothesized - underrepresented on Wikipedia, but are even \emph{slightly overrepresented} (cf. Figure \ref{fig:prop-gap}). 
Also when looking at article notable distributions of men and women, we see that \emph{articles about women tend to be longer} than articles about men (cf. Table \ref{tab:datasets_desc}) in all three datasets.
This could potentially be the result of the effort of Wikipedians to improve the coverage of minorities such as women or it can be a side product of a bias in our reference datasets which may only include very notable women, but may also cover less notable men. We addressed the later issue by selecting several reference datasets which we hope are not all subject to the same bias.



\begin{figure*}[t!]
\includegraphics[width=0.5\linewidth]{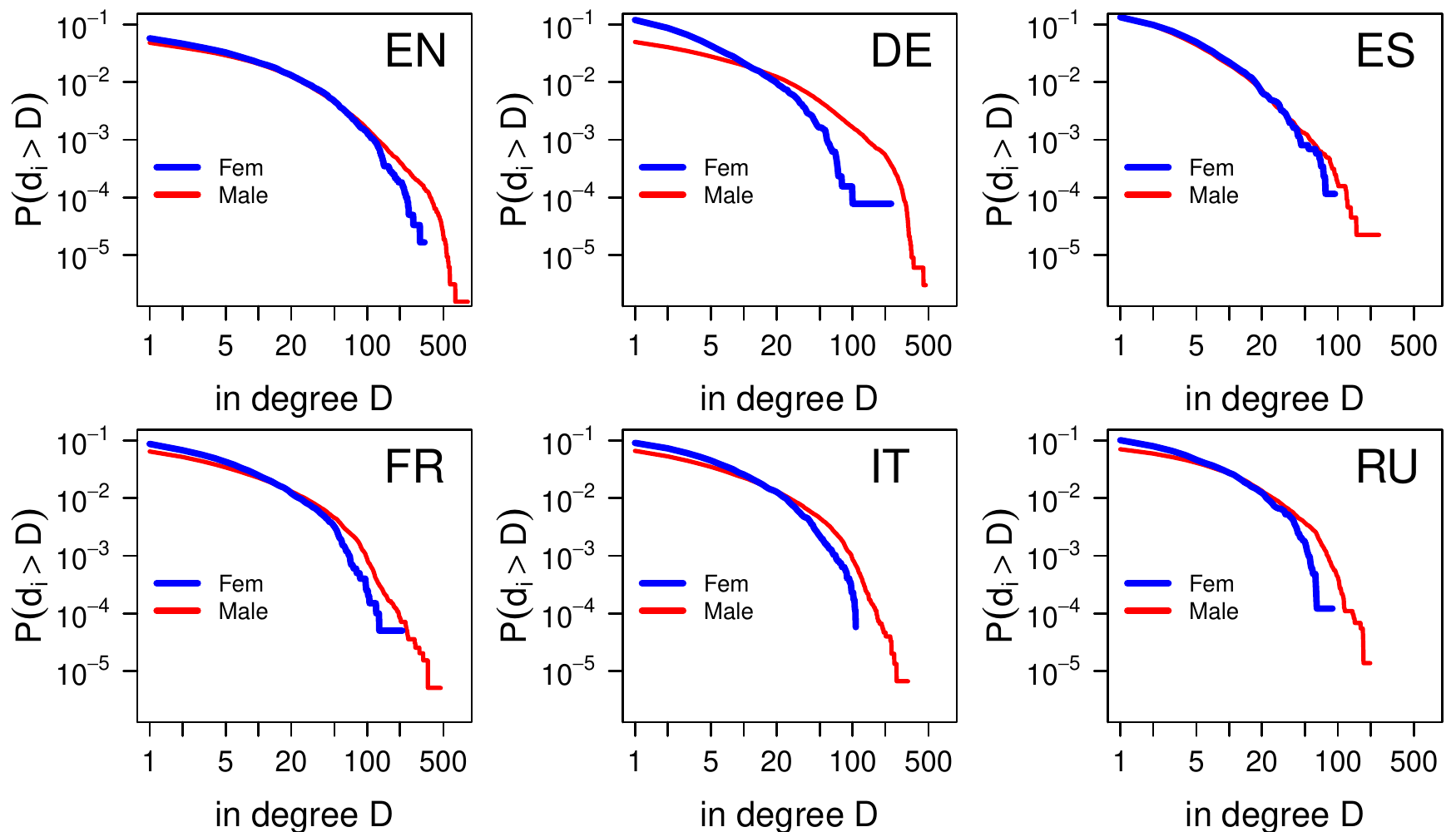}
\includegraphics[width=0.5\linewidth]{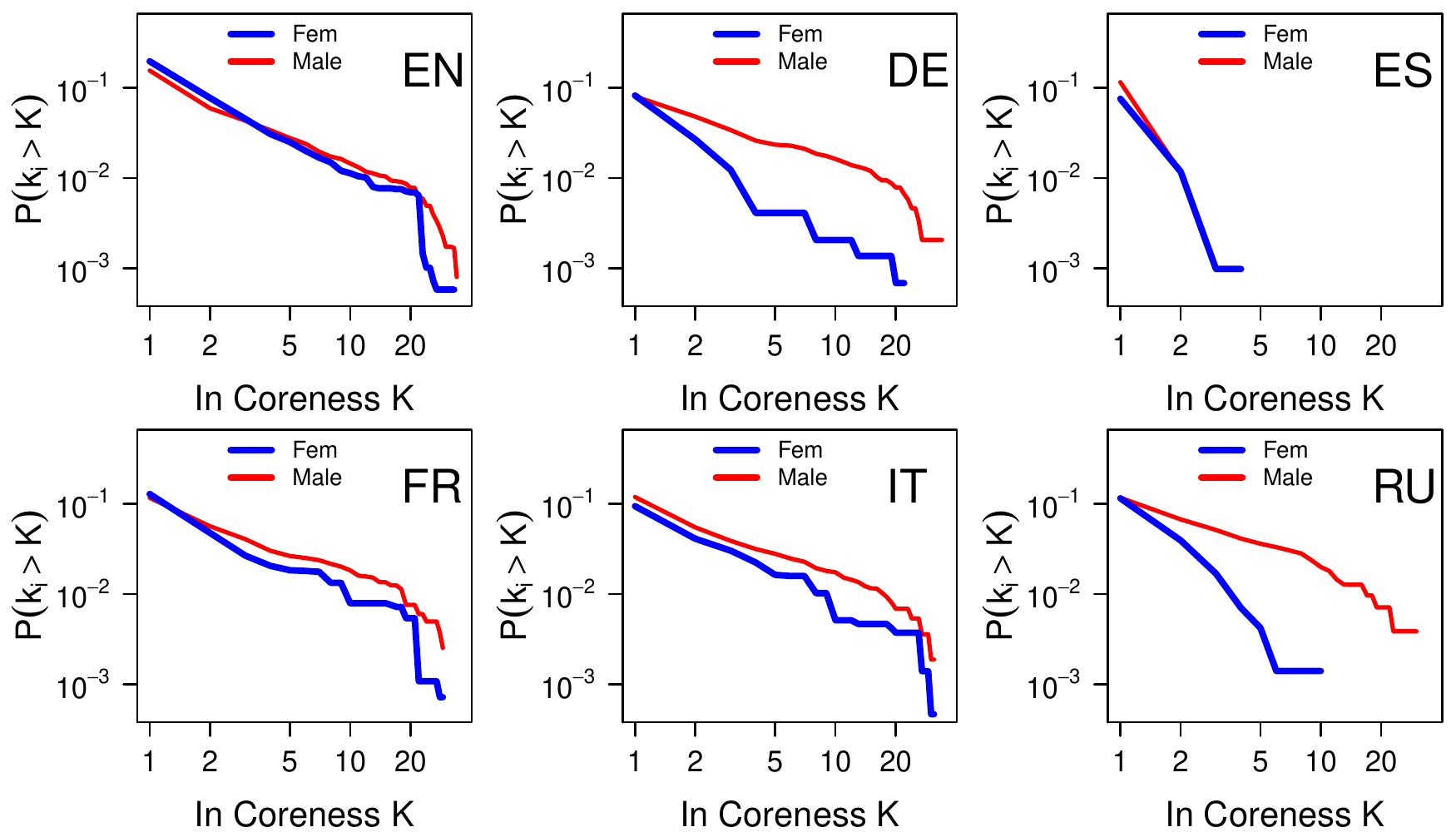}
\caption{\textbf{Structural Centrality Bias:} Complementary cumulative density function of the in-degree distributions (left) and in k-core decompositions (right)
of articles about men and women in six language editions. In some language editions like the English (EN), the Russian (RU) and the German (DE) one, men are always significantly more central than women, no matter how we measure centrality, while in others like the Spanish (ES) one, women and men are either equally central or women are more central.
}
\label{fig:centralitydists}
\end{figure*}

\subsection{Structural Bias}

Figure \ref{fig:assortMatrices} shows the logarithmic assortativity matrices of articles about men and women in six different language editions of Wikipedia based on our largest reference dataset, Freebase. The assortativity of connections within genders becomes apparent for the minority class, women, in all cases (cf. high values of $L(F,F)$). The matrices also provide a comparison across genders: $L(F,M)$ and $L(M,F)$ are both slightly negative in all language edition, which means that women connect less to men and men less to women than we would expect. 
All language editions show an asymmetry of connectivity across genders, even when we correct for overall incidence in Equation \ref{eq:L}. 
The value of $L(F,M)$ tends to be higher than $L(M,F)$, which means that men link even less to women than women to men. 

Figures \ref{fig:assort} and \ref{fig:asymmetries} show the arithmetic mean of the empirical point estimates of assortativity and asymmetry for bother gender, in comparison with the values in the three null models. It is evident that the three randomization methods destroy any kind of assortativity or asymmetry pattern, and that the empirical estimates
are significant in comparison to the narrow confidence interval of the null models. Assortativity is positive in all cases, indicating that \emph{articles about people with the same gender tend to link to each other}. For the case of asymmetry, there is a positive value of $A$ (which we defined as $A=L(F,M) - L(M,F)$) in all six language editions, validating our observation that \emph{articles about women tend to link more to  articles about men than the opposite}.

The above results show the existence of assortativity and asymmetry across genders controlling for degree.
However, structural biases can also manifest in the centrality measures, as suggested by the \emph{Smurfette principle} \cite{Pollitt1991}. 
To test the existence of this principle, we compare in-degree and k-coreness of articles about men and women on Wikipedia. 
Figure \ref{fig:centralitydists} shows the complementary cumulative density functions $P(d_i>D)$ for in-degree and 
$P(k_i>K)$ for in k-coreness in the six networks. An initial observation reveals that, in general, the tail of in-degree and in k-coreness of male articles is longer than for women articles, which is specially pronounced in the case of k-coreness of German and Russian.
We validate the above observations by measuring the distance between the two distributions and test the significance of the distance through a two-tailed Wilcoxon tests and Kolmogorov-Smirnov test (cf. Table \ref{tab:centTests}).
Our results highlight that, according to their in-degree distribution, men are indeed significantly more central in all language editions with $p<0.05$ except in the Spanish one where men and women are equally central.
The k-coreness distributions suggest that in all language editions except the Spanish, the Italian and the French one, men are more central then women.
This indicates, in some language editions like the English, the Russian and the German one, men are always significantly more central  than women, no matter how we measure centrality.


\begin{table}[b!]
\footnotesize
\centering
\begin{tabular}{c|cccccc}
    & $W_i$ & $p_i<$ & $ks_i<$ & $W_k$ & $p_k<$ & $ks_k<$\\\hline
EN  & $-$  & $10^{-15}$  & $10^{-15}$  & $-$  & $0.03$  & $10^{-4}$  \\
ES  & $+$  & $0.17$  & $0.02$  & $+$  & $10^{-4}$  & $10^{-4}$   \\
DE  & $-$  & $10^{-15}$  & $10^{-15}$  & $-$  & $10^{-12}$  & $10^{-8}$ \\
FR  & $-$  & $10^{-9}$  & $10^{-5}$  & $-$  & $0.07$  & $0.09$   \\
IT  & $-$  & $10^{-6}$  & $10^{-3}$  & $+$  & $0.95$  & $10^{-4}$  \\
RU  & $-$  & $10^{-4}$  & $10^{-7}$  & $-$  & $0.55$  & $0.003$   \\ 
\end{tabular}
\caption{\textbf{Significance of Structural Centrality Bias:}
Differences between the in-degree distributions ($W_i$)  and k-coreness distributions ($W_k$) of men and women.
A positive difference ($+$) indicates that women are more central, while a negative difference ($-$) indicates that men are more central.
The significance of the difference as suggested by the Wilcoxon test ($p_i<$) and by the Kolmogorov-Smirnov test ($ks_i<$).
In some language editions like the English (EN), the Russian (RU) and the German (DE) one, men are indeed significantly more central than women according to both centrality measures.
}\label{tab:centTests}
\end{table}

\begin{figure*}[t!]
\centering
  \subfigure[Proportion of the 150 most discriminative words for women per category]{
\includegraphics[width=1\textwidth]{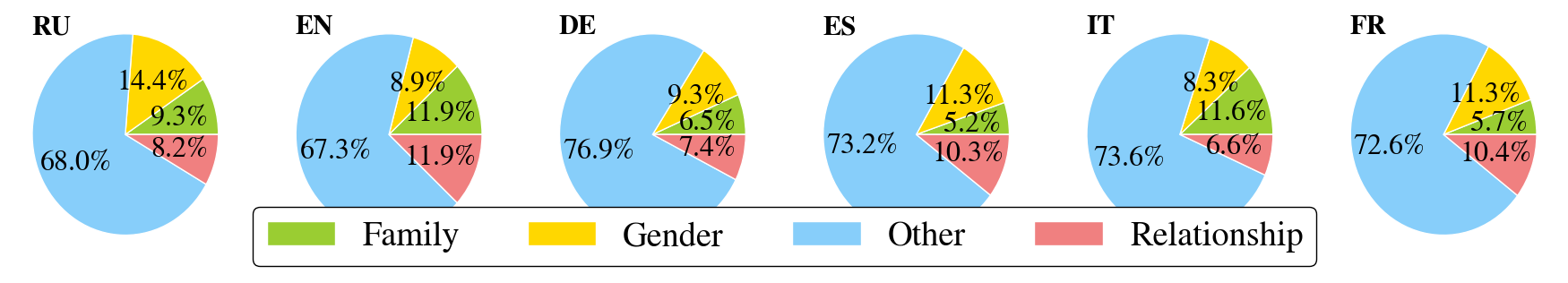}
\label{fig:cats_pies}
}
\subfigure[Proportion of the N most discriminative words for women per category]{
\includegraphics[width=1\linewidth]{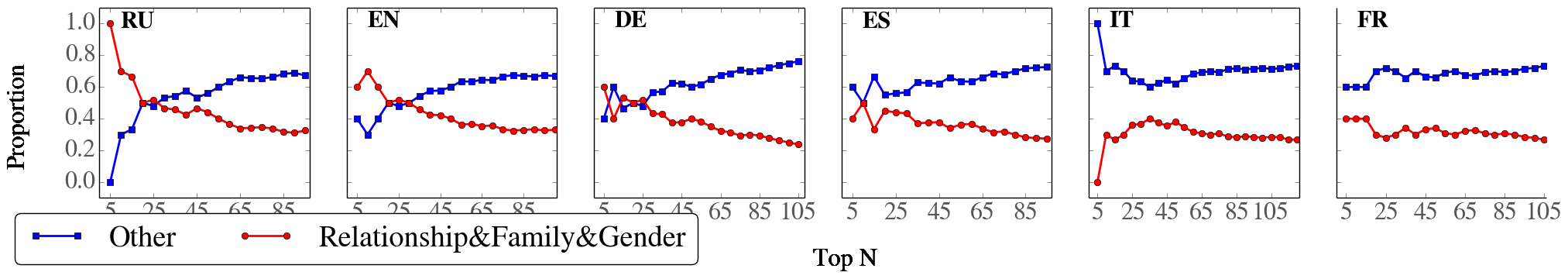}
\label{fig:cats_lines}
}
\caption{\textbf{Lexical Bias:} The proportion of the 150 most discriminative words of articles about women that belong to different categories. In all language editions between 32\% and 23\% of the 150 most indicative words for women belong to one of the three categories, while only between 0\% and 4\% of the most discriminative words for men belong to one of these categories.
In some language edition, like the Russian (RU), the English (EN) and the German (DE) one, the proportion of the most discriminative words that belong to one of these three categories is especially high among the top words. 
}
\label{fig:cats}
\end{figure*}

\subsection{Lexical Bias}
Our lexical analysis reveals that articles about women tend to emphasize the fact that they are about a women (i.e., they contain  words like ``woman'', ``female'' or ``lady''), 
while articles about men don't contain words like ``man'', ``masculine'' or ``gentleman''. The lower salience of male-related words in articles about men can be related to the concept of male as the \emph{null gender} \cite{Fox2006}, which suggests that there is a social bias to assume male as the standard gender in certain social situations. This would imply that male-defining words are not necessary because the context already defines the gender of the person the article talks about. This seems to be a plausible assumption due to the imbalance between the number of articles about men and women (cf. Table \ref{tab:datasets_desc}).


We also noticed that the relationship status and family related issues seem to be more extensively discussed in articles about woman since words like ``married'', ``divorced'', ``children'' or ``family'' are much more frequently used in articles about women. 
This confirms that \emph{men and women are indeed presented differently} on Wikipedia and that those differences go beyond what we would expect due to the history of gender inequalities - i.e., the fact  that it was more difficult for women to become famous in the past, amongst others because of unequal access to resources and the fact that the history was mainly documented through the eyes of men.
We leave the question of investigating if the lexical bias on Wikipedia reflects the lexical bias from the general media or if the Wikipedia editor community introduces an additional bias because of their narrow demographics for future work.


We use log likelihood ratios for comparing different word-gender relationships.
Not surprisingly, the most indicative words for men are often related to certain domains or fields (e.g., certain sports or professions).
For example, the most discriminative word stems for men in the English Wikipedia are ``basebal'', ``footbal'' and ``infantri'' and an article that contains a word with the stem ``basebal'' is $11.5$ times more likely to be about a man than a woman. 

For women the picture is different since among the most discriminative words for women, words like ``husband'', ``female'' and ``woman'' can be found.
To gain more insights into those difference, we use the previously introduced categories of words and manually code the words with the highest likelihood ratio for men or women.
Our results clearly show that across all language editions almost all words that fall into the category Family, Relationship or Gender, reveal a high likelihood ratio for women. 
Figure \ref{fig:cats_pies} shows that between 32\% and 23\% of the 150 most indicative words for women belong to one of the three categories. 
Note that for men only 0\% and 4\% of the most discriminative words belong to one of these categories. 
That means, words that fall into one of those categories indeed indicate that an article is about a woman which suggests that lexical gender inequalities are present on Wikipedia.
Especially, in the Russian and English Wikipedia, we can see that the majority of the 25 most discriminative words of females fall into one of those three categories (cf. Figure \ref{fig:cats_lines}).

What are these words that fall into the categories Family, Relationship or Gender and discriminate men and women?  
Table \ref{tab:likelihood-en-cat} and \ref{tab:likelihood-es-cat} show the word stems with the highest gender-specific log-likelihood ratio that belong to one of the three categories.
Almost all of them are indicative for women which means that words which are indicative for men tend not to fall into these categories.
One can further see that, for instance, in the English Wikipedia an article about a notable person that mentions that the \emph{person is divorced is $4.4$ times more likely to be about a woman} rather than a man. We observe similar results in all six language editions. For example, in the German Wikipedia an article that mentions that a person is divorced is $4.7$ times more likely about a women, in the Russian Wikipedia its $4.8$ time more likely about a woman and in the Spanish, Italian and French Wikipedia it is $4.2$ times more likely about a women.

This example shows that a lexical bias is indeed present on Wikipedia and can be observed consistently across different language editions. This result is in line with \cite{Bamman2014} who also observed that in the English Wikipedia biographies of women disproportionately focus on marriage and divorce compared to those of men.
\begin{table}[b!]
\footnotesize
\centering
\begin{tabular}{ll|ll}
Category & Term & Female   & Male  \\ \hline

Relationship   & husband & 9.2 & 1.0 \\
Gender   & female & 8.2 & 1.0 \\
Relationship   & aunt & 6.5 & 1.0 \\
Gender   & women & 6.4 & 1.0 \\
Gender   & madam & 6.1 & 1.0 \\
Gender   & woman & 5.6 & 1.0 \\
Family   & grandmoth & 5.5 & 1.0 \\
Gender   & girl & 5.3 & 1.0 \\
Gender   & mrs & 4.9 & 1.0 \\
Relationship   & divorc & 4.4 & 1.0 \\
Gender   & ladi & 4.4 & 1.0 \\
Relationship   & wed & 4.3 & 1.0 \\

Relationship   & marriag & 3.8 & 1.0 \\
Relationship   & lover & 3.8 & 1.0 \\
Family   & babi & 3.7 & 1.0 \\
Family   & sister & 3.5 & 1.0 \\
Family   & child & 3.0 & 1.0 \\
Family   & mother & 3.0 & 1.0 \\



\end{tabular} 
\caption[Gender-specific Likelihood ratio of words in the English Wikipedia]{\textbf{English Gender-specific Likelihood Ratios: } Word stems with the highest gender-specific likelihood ratio in the English Wikipedia that belong to one of the three categories (Family, Relationship and Gender).
}
\label{tab:likelihood-en-cat}
\end{table}

\begin{table}[b!]
\footnotesize
\centering
\begin{tabular}{ll|ll}
Category & Term & Female   & Male  \\ \hline

Family   & embaraz  & 9.6 & 1.0 \\
Gender   &mrs & 6.1 & 1.0 \\
Gender   & femenin & 5.3 & 1.0 \\

Gender   & madam & 4.4 & 1.0 \\
Gender   & dam & 4.4 & 1.0 \\
Family   & tia  & 4.4 & 1.0 \\
Relationship   & divorci  & 4.2 & 1.0 \\

Relationship   & bod  & 4.0 & 1.0 \\
Gender   & mujer& 3.9 & 1.0 \\
Gender   & girl & 3.9 & 1.0 \\
Gender   & lady & 3.7 & 1.0 \\

Relationship   & parej  & 3.2 & 1.0 \\
Relationship   & enamor & 3.0 & 1.0 \\
Relationship   & matrimoni  & 2.9 & 1.0 \\
Relationship   & marido & 2.7 & 1.0 \\
Relationship   & viud & 2.7 & 1.0 \\
Relationship   & amant & 2.6 & 1.0 \\
Relationship   & hereder & 2.5 & 1.0 \\
Relationship   & sexual  & 2.4 & 1.0 \\
Family   & niet  &2.3 & 1.0 \\

\end{tabular} 
\caption[Female-male likelihood ratio in the Spanish Wikipedia]{\textbf{Spanish Gender-specific Likelihood Ratios: }  Word stems with the highest gender-specific likelihood ratio in the Spanish Wikipedia that belong to one of the three categories (Family, Relationship and Gender)
}
\label{tab:likelihood-es-cat}
\end{table}




\begin{figure}[t!]
\centering
\includegraphics[width=0.4\linewidth]{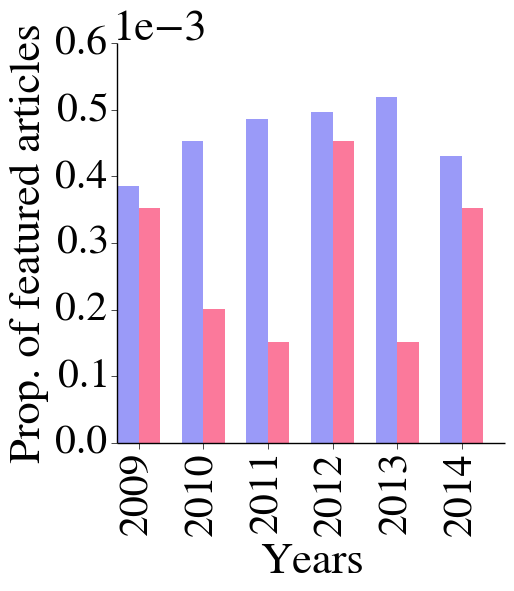}
\caption{\textbf{Visibility Bias:} The proportion of notable men and women that were featured on the front page of the English Wikipedia in the past few years. 
One can see that the proportion of men is consistently higher, but the difference is marginal.}
\label{fig:visibility}
\end{figure}

\subsection{Visibility Bias}
Figure \ref{fig:visibility} shows the proportion of notable men and women that showed up at the front page of the English Wikipedia in the past few years.
One can see that proportions of men and women that got selected are very small and therefore also the differences are marginal. 
Though we observe across all years that the proportion of men that were selected and featured at the startpage was slightly higher, the Chi-Square test suggests that the difference in proportions is not significant. Therefore, we conclude that the \emph{selection procedure of featured articles of the Wikipedia community does not suffer from gender bias}.


\section{Discussion}
\label{sec:disc}

While Wikipedia’s massive reach in coverage ensures that notable women have high likelihood of being represented on Wikipedia, evidence of gender bias surfaces from a deeper analysis of the content of those articles.
Our results clearly show that subtle lexical and structural gender biases are present on Wikipedia.

Potential explanations for these biases are the following:
it is possible that biases are a consequence of (i) the predominantly male editor community and the software design in general that might encourage male contributors and/or (ii) historic and present inequalities between men and women that manifest e.g. in unequal access to resources, unequal media presentation and historic documentation and implicit gender stereotyping (which has been shown to give men an unfair advantage in fame judgements \cite{Banaji95implicitgender}).
It seems to be plausible that certain biases such as the coverage or structural bias can be explained by historic inequalities and implicit cognitive biases due to gender stereotypes that may lead to the fact that notable men seem to be more present in our minds than notable women.
Other biases such as the lexical bias (e.g. the fact that articles about women disproportionately focus on marriage and divorce compared to articles about men) can more likely be explained by the narrow demographics of the Wikipedia editor community and the media portrayal of men and women. 
We leave the question of exploring the extent to which different factors explain different biases for future research.


\textbf{Implications}:
The low coverage and visibility bias suggest that the Wikipedia community covers notable women and men equally. However, our results highlight that editors need to pay attention to the ways women are portrayed on Wikipedia. In particular, the community needs to evaluate the gender balance of links included in articles (e.g., if an article about a woman links to the article about her husband, the husband should also link back), and to adopt a more gender-balanced vocabulary when writing articles about notable people. These existing biases might put women at a practical disadvantage: For example, because modern search and recommendation algorithms exploit both structural and textual information, women might suffer from lower visibility when it comes to ranking articles about notable people or in terms of their general visibility on Wikipedia (at least if we only take links between articles about people into account; see Figure 6 in \cite{eom2014} for preliminary comparison of ranking algorithms).


\textbf{Cross-lingual Analysis}: We observe the strongest structural bias for the English and Russian Wikipedia. Also on the lexical dimension the strongest bias becomes visible in the English and Russian Wikipedia.
Surprisingly the Spanish Wikipedia reveals the lowest structural bias. 
Comparing our results with the Gender Inequality Index of the World Economic Form (WMF) \cite{gender_inequality} shows that a positive correlation exists between the bias in the offline world and the bias on Wikipedia.
However, one needs to note that it is difficult to compare our Wikipedia based gender bias rankings of languages with the ranking of countries according to the gender inequality index since countries where the same language is predominantly spoken often reveal very different positions in the WMF ranking. 
We use the weighted average of the WMF rank positions of countries where the same language is spoken\footnote{\url{http://www.infoplease.com/ipa/A0855611.html}} and weight countries by the size of the internet population\footnote{\url{http://en.wikipedia.org/wiki/List_of_countries_by_number_of_Internet_users}}.
The Spearman rank correlation between the ranking of the 6 languages according to the WMF index shows a correlation of $0.89$ with the coverage bias based ranking, $0.37$ with the structural bias ranking and $0.09$ with the lexical bias ranking.
This indicates that to a certain extent gender inequalities of the real world manifest on Wikipedia.
However, since the Wikipedia editor community is not representative for the larger population in a country, it is also not surprising that certain biases like the lexical bias do only reveal a very limited relation with the WMF ranking.
Although Wikipedia may only reflect certain aspects of gender inequalities of the real world, gender biases that are introduced by the editor community of Wikipedia may effect the larger population and therefore it is important to investigate them.


\textbf{Reference datasets}: Our findings with regard to coverage bias are effected by the (unknown) biases inherent in the reference datasets used. Due to this, we can not make any absolute statements about coverage inequality on Wikipedia. However, regardless of this problem, 
we can assert that \emph{Wikipedia covers women and men from our reference datasets better equally well}.
Using external reference datasets that represent collections of notable people to prune down the number of biographies in Wikipedia rather than studying all of them further helps to uncouple lexical bias and structural bias from coverage bias and ensures that only people that are notable from a global perspective become the subject of study.
An alternative would be to select all people from Wikipedia using category pages such as ``Births by Year''\footnote{\url{http://en.wikipedia.org/wiki/Category:Births_by_year}} or ``Deaths by Year''\footnote{\url{http://en.wikipedia.org/wiki/Category:Deaths_by_year}} as starting point. However, these category pages do not exist in all language editions and therefore the selection would be based on the categories of the English Wikipedia only, which introduces a bias since every language editions tends to focus on their ``local heros'' \cite{Callahan2011,hecht2010tower}.



\section{Related Work}
\label{sec:relatedwork} 

\textbf{Gender Inequalities in Traditional Media:} Feminist often claim that news is not simply mostly about men, but overwhelmingly seen through the eyes of men.
In \cite{Ross2011} the authors analyze longitudinal data from the GMMP (Global Media Monitoring Project) which spans over 15 years. The authors conclude that the role of women as a producer and subject of news has seen a steady improvement, but the relative visibility of women compared to men has stuck at 1:3 which means that the world's new agencies still consider the life of men three time more worth to write about it as those of women.
Gender inequalities also manifest in films that are used for education purposes,
as revealed by the application of the Bechdel test to teaching content
\cite{Scheiner-Fisher2012}. 
In \cite{Sugimoto2013} the authors present a cross-disciplinary, global,  bibliometric analysis of the relation between gender and scientific output (i.e., number of papers, citations per paper and internationality of collaborations) using data from more than 5 million scientific publications. They find that the research output in most countries is dominated by males and that the few countries that are dominated by females have lower research output which indicates that barriers are present.

\textbf{Gender Inequalities on Wikipedia:}
Our work is not the first work which recognises the importance of understanding gender biases on Wikipedia \cite{Reagle2011,eom2014,Callahan2011,Aragon2011}.
In  \cite{Reagle2011} thousands of biographical subjects from six reference sources (e.g., The Atlantic's 100 most influential figures in American history, TIME Magazine's list of
2008's most influential people) are compared against the English-language
Wikipedia and the online Encyclopedia Britannica with respect to coverage and article length. 
The authors do not find gender-specific differences in the coverage and article length on Wikipedia, but Wikipedia’s missing articles are disproportionately female relative to those of Britannica.
Our findings on the coverage dimension confirm their findings and further we also analyze the content of articles on Wikipedia which they left for future work.

In \cite{Bamman2014} the authors present a method to learn biographical structures from text and observe that in the English Wikipedia biographies of women disproportionately focus on marriage and divorce compared to those of men, which is in line with our findings on the lexical dimension.
Recent research showed that most important historical figures across Wikipedia language
editions are born in Western countries after the 17th century, and are male \cite{eom2014}. On average only 5.2 female historic figures are observed among the top 100 persons.
The authors use different link-based ranking algorithms and focus on the top 100 figures in each language edition.
Their results clearly show that very few women are among the top 100 figures in all language editions, but since the authors do not use any external reference lists it remains unclear how many women we would expect to see among the top 100 figures.

Previous research has also explored gender inequalities in the editor community of Wikipedia and potential reasons for it (cf. \cite{LamUDSMTR11,Collier2012,Hill2013}).
Also among Wikipedians, the importance of this issue has been acknowledge for example through the initiation of the ``Countering Systemic Bias'' WikiProject\footnote{\url{http://en.wikipedia.org/wiki/Wikipedia:WikiProject_Countering_systemic_bias}} in 2004.


\textbf{Gender inequalities in Social Media:}
In \cite{Szell2013} the author study a communication network in a MMOG and find a  similar effect as \cite{Smoreda_SPQ2000}. 
Female players send about 25\% more messages (0.74 per day) than males (0.60 per day). 
Consequently, females show a significantly higher average degree in their communication networks, however, the communication partners of females have a significantly lower average degree than those of males, i.e. females have more communication partners, while males tend to have better connected ones.
Recent research \cite{MagnoWeber-socinfo2014} suggests that in Twitter and Google+ online inequality is strongly correlated to offline inequality,
but the directionality can be counter-intuitive.
In particular, they consistently observe women to have a
higher online status, as defined by a variety of measures,
compared to men in countries such as Pakistan or Egypt,
which have one of the highest measured gender inequalities.
In \cite{Garcia2014} the authors show that subconscious biases which contribute to the creation of inequality are not only present in movie scripts but also in Twitter conversations. 
Also the viewing and sharing patterns of youtube videos reveal differences in which content is consumed and discussed by different genders \cite{Abisheva2014}. This kind of differences also manifest in wall discussions in MySpace, where emotional expression patterns differ across genders \cite{Thelwall2009}.


\section{Conclusions}
\label{sec:concl}

Wikipedia seems to have successfully established processes that ensure that notable women have a high likelihood of being portrayed on Wikipedia. At the same time, our work surfaces evidence of more subtle forms of gender inequality. In particular, women on Wikipedia tend to be more linked to men than vice versa, which can put women at a disadvantage in terms of - for example - visibility or reachability on Wikipedia. In addition, we find that womens' romantic relationships and family-related issues are much more frequently discussed in their Wikipedia articles than in mens' articles. This suggests that there are gender differences w.r.t. how the Wikipedia community conceptualizes notable men/women. 
Because modern search and recommendation algorithms exploit both, structure and content, women may suffer from lower visibility in social networks (or article networks) where men (or articles about men) are more central and include more links to other men than to other women.
To reduce such effects, the editor community needs to evaluate the gender balance of links included in articles (e.g., if an article about a woman links to the article about her husband, the husband should also link back), and to adopt a more gender-balanced vocabulary when writing articles about notable people.
Further, engineers and researchers need to develop a deeper understanding of how different types of search and recommendation algorithms impact the visibility of minorities. 

In summary, the contributions of this work are twofold: (i) we present a computational method for assessing gender bias on Wikipedia \emph{along multiple dimensions} and (ii) we apply this method to several language editions of Wikipedia and share empirical insights on observed gender inequalities. We translate our findings into some potential actions for the Wikipedia editor community to reduce gender biases in the future. We hope our work contributes to increasing awareness about gender biases online, and in particular to raising attention to the different levels in which these biases can manifest themselves. The methods presented in this work can be used to assess, monitor and evaluate these issues on Wikipedia on an ongoing basis.


\small
\bibliographystyle{aaai}
\bibliography{sample}

\begin{thebibliography}{}

\bibitem[\protect\citeauthoryear{Abisheva \bgroup et al\mbox.\egroup
  }{2014}]{Abisheva2014}
Abisheva, A.; Garimella, V. R.~K.; Garcia, D.; and Weber, I.
\newblock 2014.
\newblock {Who Watches ( and Shares ) What on YouTube ? And When ? Using
  Twitter to Understand YouTube Viewership}.
\newblock In {\em WSDM'14}.

\bibitem[\protect\citeauthoryear{Amy~Yu and Hidalgo}{2013}]{Pantheon}
Amy~Yu, Kevin~Hu, S. R. D.~G., and Hidalgo, C.~A.
\newblock 2013.
\newblock {The Pantheon Multilingual Wikipedia Expression Dataset}.
\newblock MIT project.

\bibitem[\protect\citeauthoryear{Arag{\'{o}}n \bgroup et al\mbox.\egroup
  }{2012}]{Aragon2011}
Arag{\'{o}}n, P.; Kaltenbrunner, A.; Laniado, D.; and Volkovich, Y.
\newblock 2012.
\newblock Biographical social networks on wikipedia - {A} cross-cultural study
  of links that made history.
\newblock {\em Proceedings of WikiSym}.

\bibitem[\protect\citeauthoryear{Callahan and
  Herring}{2011a}]{callahan2011cultural}
Callahan, E., and Herring, S.~C.
\newblock 2011a.
\newblock Cultural bias in wikipedia content on famous persons.
\newblock {\em Journal of the Association for Information Science and
  Technology} 62(10):1899--1915.

\bibitem[\protect\citeauthoryear{Callahan and Herring}{2011b}]{Callahan2011}
Callahan, E.~S., and Herring, S.~C.
\newblock 2011b.
\newblock Cultural bias in wikipedia content on famous persons.
\newblock {\em Journal of the American Society for Information Science and
  Technology} 62(10):1899--1915.

\bibitem[\protect\citeauthoryear{Collier and Bear}{2012}]{Collier2012}
Collier, B., and Bear, J.
\newblock 2012.
\newblock Conflict, criticism, or confidence: An empirical examination of the
  gender gap in wikipedia contributions.
\newblock In {\em Proceedings of the ACM 2012 Conference on Computer Supported
  Cooperative Work}, CSCW '12,  383--392.
\newblock New York, NY, USA: ACM.

\bibitem[\protect\citeauthoryear{Eom \bgroup et al\mbox.\egroup
  }{2014}]{eom2014}
Eom, Y.-H.; Arag{\'{o}}n, P.; Laniado, D.; Kaltenbrunner, A.; Gigna, S.; and
  Shepelyansky, D.~L.
\newblock 2014.
\newblock Interactions of culture and top people of wikipedia from ranking 24
  language editions.
\newblock {\em Plos One}.

\bibitem[\protect\citeauthoryear{Finkbeiner}{2013}]{Finkbeiner}
Finkbeiner, A.
\newblock 2013.
\newblock {What I’m not going to do: Do media have to talk about family
  matters?}
\newblock DoubleXScience.

\bibitem[\protect\citeauthoryear{Fox, Johnson, and Rosser}{2006}]{Fox2006}
Fox, M.~F.; Johnson, D.~G.; and Rosser, S.~V.
\newblock 2006.
\newblock {\em Women, gender, and technology}.
\newblock University of Illinois Press.

\bibitem[\protect\citeauthoryear{Garcia, Weber, and
  Garimella}{2014}]{Garcia2014}
Garcia, D.; Weber, I.; and Garimella, V. R.~K.
\newblock 2014.
\newblock Gender asymmetries in reality and fiction: The bechdel test of social
  media.
\newblock In {\em International AAAI Conference on Weblogs and Social Media
  (ICWSM)}.

\bibitem[\protect\citeauthoryear{Giatsidis, Thilikos, and
  Vazirgiannis}{2013}]{Giatsidis2013}
Giatsidis, C.; Thilikos, D.~M.; and Vazirgiannis, M.
\newblock 2013.
\newblock D-cores: measuring collaboration of directed graphs based on
  degeneracy.
\newblock {\em Knowledge and information systems} 35(2):311--343.

\bibitem[\protect\citeauthoryear{Goodman and Vertesi}{2012}]{Goodman2012}
Goodman, E., and Vertesi, J.
\newblock 2012.
\newblock Design for x?: distribution choices and ethical design.
\newblock In {\em CHI'12 Extended Abstracts on Human Factors in Computing
  Systems},  81--90.
\newblock ACM.

\bibitem[\protect\citeauthoryear{Hecht and Gergle}{2010}]{hecht2010tower}
Hecht, B., and Gergle, D.
\newblock 2010.
\newblock The tower of babel meets web 2.0: user-generated content and its
  applications in a multilingual context.
\newblock In {\em Conference on Human Factors in Computing Systems},  291--300.
\newblock ACM.

\bibitem[\protect\citeauthoryear{Hill and Shaw}{}]{Hill2013}
Hill, B.~M., and Shaw, A.
\newblock {\em PLoS One}.

\bibitem[\protect\citeauthoryear{Lam \bgroup et al\mbox.\egroup
  }{2011}]{LamUDSMTR11}
Lam, S.~K.; Uduwage, A.; Dong, Z.; Sen, S.; Musicant, D.~R.; Terveen, L.~G.;
  and Riedl, J.
\newblock 2011.
\newblock Wp: clubhouse?: an exploration of wikipedia's gender imbalance.
\newblock In Ortega, F., and Forte, A., eds., {\em Int. Sym. Wikis},  1--10.
\newblock ACM.

\bibitem[\protect\citeauthoryear{Magno and
  Weber}{2014}]{MagnoWeber-socinfo2014}
Magno, G., and Weber, I.
\newblock 2014.
\newblock International gender differences and gaps in online social networks.
\newblock In {\em The 6th International Conference on Social Informatics
  (SocInfo)}.

\bibitem[\protect\citeauthoryear{Murray}{2003}]{Murray2003}
Murray, C.
\newblock 2003.
\newblock {\em Human Accomplishment. The pursuit of excellence in the Arts and
  Sciencesy}.

\bibitem[\protect\citeauthoryear{Newman}{2003}]{Newman2003}
Newman, M. E.~J.
\newblock 2003.
\newblock Mixing patterns in networks.
\newblock {\em Phys. Rev. E} 67(2):026126.

\bibitem[\protect\citeauthoryear{Pollitt}{1991}]{Pollitt1991}
Pollitt, K.
\newblock 1991.
\newblock {Hers; The Smurfette Principle}.
\newblock The New York Times.

\bibitem[\protect\citeauthoryear{Reagle and Rhue}{2011}]{Reagle2011}
Reagle, J., and Rhue, L.
\newblock 2011.
\newblock Gender bias in {Wikipedia} and {Britannica}.
\newblock {\em International Journal of Communication} 5.

\bibitem[\protect\citeauthoryear{Ross and Carter}{2011}]{Ross2011}
Ross, K., and Carter, C.
\newblock 2011.
\newblock {Women and news: A long and winding road}.
\newblock {\em Media, Culture and Society} 33(8):1148--1165.

\bibitem[\protect\citeauthoryear{Scheiner-Fisher and
  Russell}{2012}]{Scheiner-Fisher2012}
Scheiner-Fisher, C., and Russell, W.~B.
\newblock 2012.
\newblock {Using Historical Films to Promote Gender Equity in the History
  Curriculum}.
\newblock {\em The Social Studies} 103(6):221--225.

\bibitem[\protect\citeauthoryear{Schich \bgroup et al\mbox.\egroup
  }{2014}]{Schich2014}
Schich, M.; Song, C.; Ahn, Y.-Y.; Mirsky, A.; Martino, M.; Barab\'{a}si, A.-L.;
  and Helbing, D.
\newblock 2014.
\newblock A network framework of cultural history.
\newblock {\em Science} 345(6196):558--562.

\bibitem[\protect\citeauthoryear{Schwab \bgroup et al\mbox.\egroup
  }{2013}]{gender_inequality}
Schwab, K.; Brende, B.; Zahidi, S.; Bekhouche, Y.; Guinault, A.; Soo, A.;
  Hausmann, R.; and Tyson, L.~D.
\newblock 2013.
\newblock {The Global Gender Gap Report 2013}.
\newblock World Economic Forum.

\bibitem[\protect\citeauthoryear{Schwartz \bgroup et al\mbox.\egroup
  }{2013}]{Schwartz2013}
Schwartz, H.~A.; Eichstaedt, J.~C.; Kern, M.~L.; Dziurzynski, L.; Ramones,
  S.~M.; Agrawal, M.; Shah, A.; Kosinski, M.; Stillwell, D.; Seligman, M.~E.;
  and Ungar, L.~H.
\newblock 2013.
\newblock Personality, gender, and age in the language of social media: the
  open-vocabulary approach.
\newblock {\em PLoS One} 8(9).

\bibitem[\protect\citeauthoryear{Smoreda and Licoppe}{2000}]{Smoreda_SPQ2000}
Smoreda, Z., and Licoppe, C.
\newblock 2000.
\newblock {Gender-Specific Use of the Domestic Telephone}.
\newblock {\em Social Psychology Quarterly} 63(3):238--252.

\bibitem[\protect\citeauthoryear{Sugimoto \bgroup et al\mbox.\egroup
  }{2013}]{Sugimoto2013}
Sugimoto, Cassidy R.and~Larivi\`{e}re, V.; Ni, C.; Gingras, Y.; and Cronin, B.
\newblock 2013.
\newblock {Bibliometrics: Global gender disparities in science}.
\newblock {\em Nature} 504(7479):211--213.

\bibitem[\protect\citeauthoryear{Szell and Thurner}{2013}]{Szell2013}
Szell, M., and Thurner, S.
\newblock 2013.
\newblock How women organize social networks different from men.
\newblock {\em Scientific Reports} 3(1214).

\bibitem[\protect\citeauthoryear{Thelwall, Wilkinson, and
  Uppal}{2009}]{Thelwall2009}
Thelwall, M.; Wilkinson, D.; and Uppal, S.
\newblock 2009.
\newblock {Data mining emotion in social network communication: Gender
  differences in MySpace}.
\newblock {\em Journal of the American Society for Information Science and
  Technology} 61(1):1--10.

\bibitem[\protect\citeauthoryear{Thomas}{1992}]{Thomas1992}
Thomas, G.
\newblock 1992.
\newblock {\em A position to command respect : women and the eleventh
  Britannica}.
\newblock Scarecrow Press.

\end{thebibliography}

\end{document}